# The PICARD Payload Data Centre

G. Pradels[1], T. Guinle[2]
*CNES, Toulouse, 31401, France*

G. Thuillier[3], A. Irbah[4], J-P. Marcovici[5], C. Dufour[6]
*CNRS/SA, Verrières le Buisson, 91371, France*

D. Moreau[7], C. Noel[8], M. Dominique[9]
*BUSOC, Uccle, 1180, Belgium*

T. Corbard[10], M. Hadjara[11]
*UNSA/CNRS, Nice, 06304, France*

S. Mekaoui[12]
*RMIB, Uccle ,1180, Belgium*

C. Wehrli[13]
*PMOD/WRC, Davos, 7260, Switzerland*

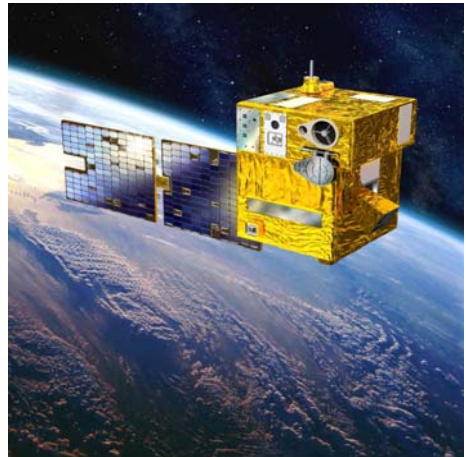

**PICARD is a scientific space mission dedicated to the study of the solar variability origin. A French micro-satellite will carry an imaging telescope for measuring the solar diameter, limb shape and solar oscillations, and two radiometers for measuring the total solar irradiance and the irradiance in five spectral domains, from ultraviolet to infrared. The mission is planed to be launched in 2009 for a 3-year duration. This article presents the PICARD Payload Data Centre, which role is to collect, process and distribute the PICARD data. The Payload Data Centre is a joint project between laboratories, space agency and industries. The Belgian scientific policy office funds the industrial development and future operations under the European Space Agency program. The development is achieved by the SPACEBEL Company. The Belgian operation centre is in charge of operating the PICARD Payload Data Centre. The French space agency leads the development in partnership with the French scientific research centre, which is responsible for providing all the scientific algorithms. The architecture of the PICARD Payload Data Centre (software and hardware) is presented. The software system is based**

---

[1] Ground segment engineer, CNES/DCT/PS/CMI, gregory.pradels@cnes.fr, 18 Av. Edouard Belin
[2] Ground segment engineer, CNES/DCT/PS/CMI, thierry.guinle@cnes.fr, 18 Av. Edouard Belin
[3] PI of the PICARD mission, CNRS/SA, gerard.thuillier@aerov.jussieu.fr, Rte des Gâtines 91371
[4] Data processing engineer, CNRS/SA, abdenour.irbah@aerov.jussieu.fr, Rte des Gâtines 91371
[5] Embedded software engineer, CNRS/SA, jean-pierre.marcovici@aerov.jussieu.fr, Rte des Gâtines 91371
[6] System engineer, CNRS/SA, christophe.dufour@aerov.jussieu.fr, Rte des Gâtines 91371
[7] Manager, BUSOC, didier.moreau@busoc.be, 3 Avenue Circulaire
[8] System engineer, BUSOC, christian.noel@busoc.be, 3 Avenue Circulaire
[9] Data processing engineer, BUSOC, marie.dominique@busoc.be, 3 Avenue Circulaire
[10] Astronomer, UNSA/CNRS, thierry.corbard@oca.eu, Blvd de l'observatoire 06304
[11] Data processing engineer, UNSA/CNRS, massinissa.hadjara@obs-azur.fr, Blvd de l'observatoire 06304
[12] Data processing engineer, RMIB, sabri.mekaoui@oma.be, Ringlaan 3 1180
[13] Data processing engineer, PMOD/WRC, christoph.wehrli@pmodwrc.ch, Dorfstrasse 7260

**on a Service Oriented Architecture. The host structure is made up of the basic functions such as data management, task scheduling and system supervision including a graphical interface used by the operator to interact with the system. The other functions are mission-specific: data exchange (acquisition, distribution), data processing (scientific and non-scientific processing) and managing the payload (programming, monitoring). The PICARD Payload Data Centre is planned to be operated for 5 years. All the data will be stored into a specific data centre after this period.**

## I. Introduction to the PICARD mission

### A. Scientific rationale

During the XVII$^{th}$ century, the gravitation was the dominant topic within the European astronomers. This is why, the determination of the planet orbit was important. Kepler measured the Mars orbit, but his method was not applicable to the Earth. Jean Picard (1620-1682), a French astronomer had the bright idea to measure the solar diameter as a function of the day of the year. The diameter variation has allowed derivation of the Earth's orbit eccentricity. After his death, this program was continued by Philippe de la Hire allowing gathering data during 70 years. Fortunately, the observations cover the Maunder minimum, a period showing a quasi absence of sunspots, till the Sun resumed its activity by 1715. The data were re-analysed by Ribes [1] who found that the Sun diameter was larger during that minimum than when the Sun was active. Present solar models indicate a solar luminosity smaller during the Maunder Minimum than nowadays. Consequently, an anticorrelation between luminosity and solar diameter was suggested by these measurements. This has been extensively discussed in terms of quality of measurements and theoretical modelling.

From the ground, different techniques are used as astrolabes, Mercury transits in front the Sun, solar eclipses, meridian measurements. All of them show inconsistent results with correlation between solar diameter with activity, or anticorrelation or no variation at all (Thuillier [2]). At this point, it should be kept in mind that no consensus in terms of spectral domain of measurements, instrument design and data processing was agreed. The instrumental and atmospheric effects can explain most of the discrepancies. Outside the atmosphere, few results exist: some observations from stratospheric balloons (Djafer [3]), and from the MDI instrument (Kuhn [4]) on board the Solar and Heliospheric Observatory (SoHO) satellite.

There are important solar parameters that are key constraints for validating the physics of solar interior models, namely:
- Solar diameter, limb shape, asphericity in the photosphere,
- Total solar irradiance (TSI),
- Oscillation modes,
- Temperature,
- Solar spectrum.

and their variability as a function of solar activity. Among these parameters, the solar diameter presents the greatest uncertainty. The importance of the possible variation of the photospheric diameter with solar activity arises from theoretical studies of the convective zone physics and in particular the role of the turbulence within the solar plasma (Sofia [5]). Consequently PICARD aims at simultaneously measuring several of the above parameters from space to achieve the following scientific objectives:
- Modelling of the solar machine using simultaneous measurements of several fundamental parameters and knowing their variability. In particular, it will be investigated the role of the magnetic field, on surface or deeper in the convective zone to elucidate the origin of the solar activity.
- Contribution to solar luminosity reconstruction for use in climate models.
- Long terms trend using the solar diameter referred to stars angular distances.
- Understanding of the ground based measurements by comparing them to in orbit measurements.

Strong synergies with Solar Dynamics Observer, the SOLAR mission on board the International Space Station, and ground based observatories are foreseen.

### B. The measurements of the PICARD mission

The following measurements will be carried out in orbit:
- Diameter, limb shape and asphericity in the photospheric continuum at 535, 607 and 782 nm, and at 215 nm to study the chromosphere influence on the diameter. A precision of 3 mas per single image is expected. The diameters will be referred to stars angular distances.
- Solar activity will be measured at 215 nm and within the Ca II line (393 nm).
- Five solar spectral channels will be observed at the same wavelengths as above.
- TSI will be measured by two independent radiometers as on SoHO.
- Solar oscillations at 535 nm on the solar limb and using macropixels.

The variability of the above quantities are key inputs for ht solar modelling. This is why the launch is foreseen at spring of 2009 to benefit of the solar activity rising. The following measurements will be carried out on the ground:
- Diameter, limb shape and asphericity by the same instrument as in orbit,
- Local atmospheric turbulence,
- Solar images at 393 nm and 607 nm as supporting the measurement in orbit.

Limb shape, diameter asphericity may depend on the wavelength. This is why measurements are carried out as a function of wavelength. Measurements at 215 nm and 393 nm will contribute to detect limbs at active regions (sunspots and faculae), which may affect the diameter determination.

### C. The instruments of the PICARD mission

In orbit two different types of radiometers will be used to discriminate between variations of instrumental origin and of solar origin: SOlar VAriability PICARD (SOVAP) provided by the Institut Royale de Météorologie Belge (IRM-B) and PREcision MOnitoring Sensors (PREMOS) provided by the Physikalisch-Metorologisches Observatorium Davos (PMOD). They use the same configuration as on SoHO with the IRM-B radiometer and the PMOD photometer. The third instrument, SOlar Diameter Imager and Surface Mapper (SODISM), is a Cassegrain telescope associated to a 2k x 2k Charge-Coupled Device (CCD) detector. It is provided by the Service d'Aéronomie (SA) of the Centre National de Recherche Scientifique (CNRS). The instrument has two filters wheels carrying the interference filters selecting the spectral domains of measurements. Furthermore, the instrument has an entrance window to limit the solar energy input, and avoiding unuseful warming. Four prisms generate secondary images from the sun located in the four CCD corners (Assus [13]). They allow monitoring of the instrument geometrical scale, i. e. the relationship between pixel distance and angle. From time to time, by returning the spacecraft, doublets of stars having an angular separation comparable to the solar diameter are measured for absolute instrument calibration.

**Figure 1. PICARD Payload**

SOVAP consists in a radiometer and a bolometer provided by IRM (Belgian). PREMOS consists in a radiometer and three sunphotometers under responsibility of PMOD (Switzerland). SODISM is a metrological imaging telescope under responsibility of CNRS/SA (France).

For these three instruments electronics achieve data handling, formatting, sequencing and thermal regulation.

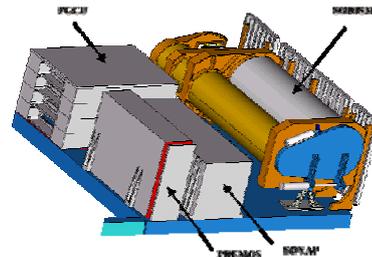

The sunphotometers are basically made of a detector, an interference filter, optics and a shutter. The radiometer is made of a cavity receiving the solar energy, and another cavity hidden from the sun, is warmed by Joule effect. When the two cavities are in thermal equilibrium, the energy provided to the hidden cavity measures the energy received from the sun.

At Plateau de Calern, the same instrument as SODISM will measure the solar diameter and limb shape. The Moniteur d'Images Solaires Franco-Algérien (MISolFA) measures the local turbulence. Collaboration with on-ground observatories is foreseen.

The scientific objectives and instruments are described in Thuillier [6].

### D. Cooperation for the development of the PICARD Satellite and the PICARD Payload Data Centre

The PICARD space mission is led by the Centre National d'Etudes Spatiales (CNES), France, in partnership with CNRS/SA. CNES handles the overall project and provides the most important part of facilities (satellite bus, Earth terminals, control centre, …), funding and human resources. CNRS leads the science team and conducts the development of the payload. Nevertheless, as usual in space activities, international cooperation is a key input. At science level, PICARD team gathers scientists from France, Belgium, Canada, Switzerland and United States. The payload is built by laboratories from France, Belgium and Switzerland. At ground level, the cooperation concerns the PICARD Payload Data Centre, which is dedicated to load commands to the instruments, to receive, process and distribute data.

Since the beginning of the PICARD project, the BELgian Scientific POlicy office (BELSPO) has been interested in PICARD Payload Data Centre with the objective to operate it at the Belgian User Support And Operation Centre (B-USOC) located at Uccle in Brussels. Thus, PICARD Payload Data Centre development, validation and operations are and will be achieved thanks to the Belgium contribution. CNES has organised the different project phases and activities in order to optimize this cooperation. As Belgium is member of the PROgramme de Développement d'EXpériences scientifiques (PRODEX) from the European Space Agency (ESA), an Agreement has been established between CNES, ESA, and Belgium.

Figure 2 depicts the project organisation. Definition, development and validation of the PICARD Payload Data Centre is under CNES responsibility. An integrated team composed by CNES and B-USOC, with support from CNRS/SA, has produced the requirements. The realisation is currently handled by the Belgian subcontractor SPACEBEL. Scientific algorithms are defined and developed by laboratories of CNRS, IRMB and PMOD. CNRS/SA collects all the software components and is in charge of deliveries to CNES for integration into PICARD Payload Data Centre by the subcontractor. The latter will be located and operated at the B-USOC. The end of the PICARD commissioning phase (3 months after launch) is an important milestone for PICARD Payload Data Centre. At that time, the overall responsibility will be transferred from CNES to B-USOC for the all mission lifetime.

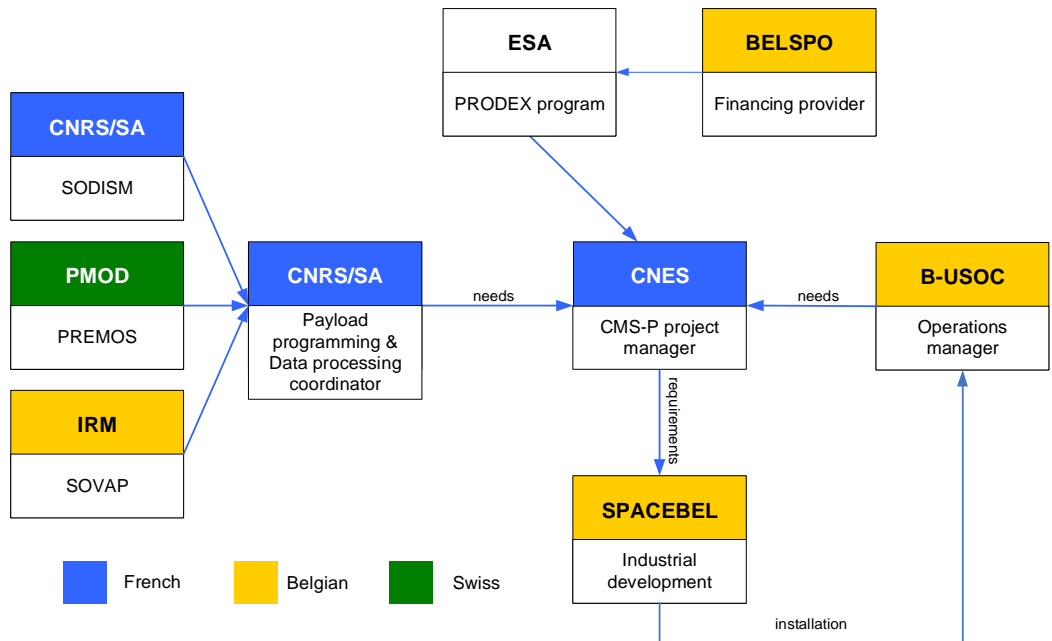

**Figure 2. Cooperation for the development of the PICARD Payload Data Centre**

## II. Architecture of a MYRIADE mission

**A. The CNES small scientific mission concepts**

The CNES small satellite programs provides the laboratories with a possibility to fly scientific experiments on board small satellites (Tatry [7]). The CNES MYRIADE program is dedicated to small scientific and technological space missions This program has been used for several missions (http://smsc.cnes.Fr/HomeFr.html):

- Detection of Electro-Magnetic Emissions Transmitted from Earthquake Regions (DEMETER) has been launched in 2004 to study the ionospheric disturbances in relation to the seismic activity and to the human activity (Lagoutte [9])
- Essaim is a military constellation of 4 satellites for testing satellite ability to collect communications intelligence in several frequency bands from strategically interesting target areas. The satellites have launched in 2004.
- Polarization and Directionality of the Earth's Reflectances (PARASOL) has been launched in 2004, with Essaim, for measuring the directional characteristics and the polarization of the light reflected by the Earth/atmosphere couple (Lier [10])

and soon for:

- PICARD is a solar characterisation mission (see I). It is indented to be launched in 2009 (Buisson [9]),
- MICRO-Satellite à traînée Compensée pour l'Observation du Principe d'Equivalence (MICROSCOPE) is a mission of fundamental physics (test of the Equivalence Principle). The launch is foreseen to 2011 (Dubois [11])
- Tool for the Analysis of RAdiations from lightNIngs and Sprites (TARANIS) is a mission dedicated to the study of the coupling between atmosphere, ionosphere and magnetosphere. It is foreseen to be launched in 2001 (Bastien-Thiry [12]).

The main objective of this program is to facilitate access to space to experimenters with reduced delays and cost. CNES deals with the satellite building while laboratories deal with the scientific payload and the Payload Data Centre. After the first MYRIADE missions, the construction of the satellite platform and of the satellite control centre is mature. The payload remains the critical element due to the new technology involved to achieve the scientific goal of the mission. In order to help laboratories handling the development of the data centre and the payload, CNES proposes its own expertise and resources (Pradels [8]).

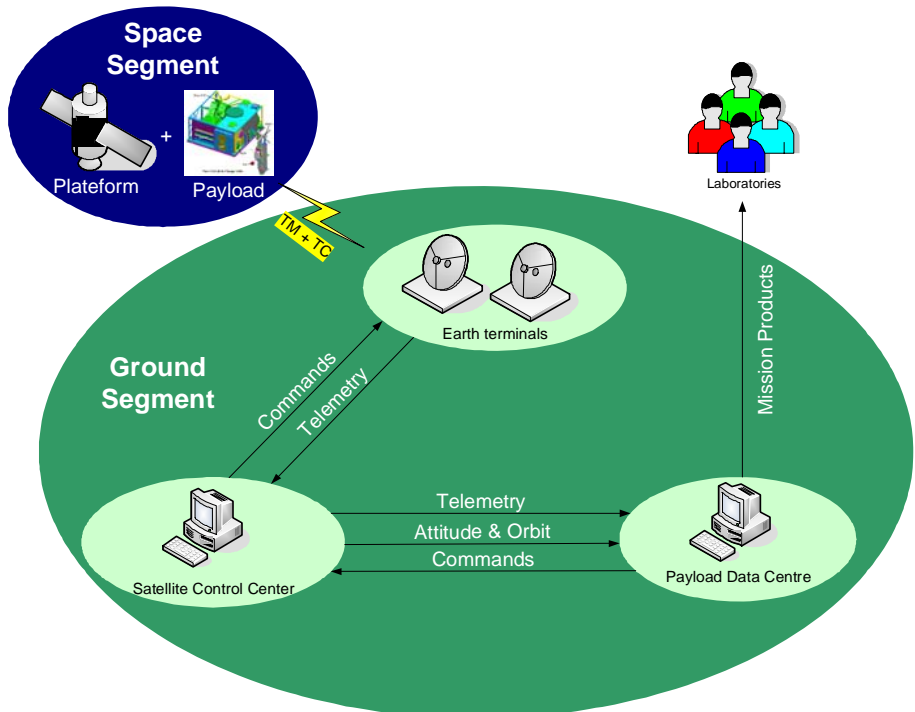

**Figure 3. Architecture of a CNES small scientific space mission**

Figure 3 illustrates the contents of the space and ground segment for a MYRIADE mission. The role of each segment can be summarized as follows: the space segment communicates with the ground segment via a network of stations. The Satellite Control Centre acquires the telemetry, handles the satellite operations and provides to the Payload Data Centre the data necessary for the payload operations and scientific processing. The Payload Data Centre collects the mission data, handles the payload operations and delivers the mission products to the laboratories.

### B. The Satellite Platform
MYRIADE is a versatile satellite platform dedicated to small scientific or technological missions. It can accept payloads between 100 and 150 kg. The platform is equipped with independent elements, which can be configured, changed or modified to fit with each mission.

### C. Earth Terminal
The Satellite Control Centre uses a network of 2 GHz Earth terminal to send commands. They are placed all around the world: Aussaguel (France), Kiruna (Sweden) but also South of Africa, French Guyana and Kerguelen Island. These terminals are directly operated from CNES facilities in Toulouse. The network allows the Satellite Control Centre to receive data from several satellites.

### D. The Satellite Control Centre
The multi-satellite control centre is able to simultaneously control seven satellites out of five different missions. It is divided into several software units:
- G1 in charge of satellite/ground interfaces management,
- G2 in charge of orbits determination and attitude control,
- G3 in charge of telemetry storage and offline processing.
- G4 in charge of interfacing the Payload Mission Centre.
- A Data Remote Processing PC (DRPPC) allowing to visualize or to monitor the telemetry,
- A WWW server allowing the DRPPC to retrieve data from the Satellite Control Centre,
- A task scheduler in charge of the Satellite Control Centre automation,
- SYGALE in charge of the alarms management.

### E. The Payload Data Centre
The Payload Data Centre is the last element of the operational processing sequence and also the interface between the payload and the mission users. It is the exploitation centre of the mission (Pradels [12]). It has the following functions:
- Payload control: the commands necessary to control the payload are generated at least once a week. The command plan is sent to the Satellite Control Centre, which uplinks it directly to the satellite. This function is thus very critical.

- Data acquisition: a Payload Data Centre collects various types of data such as telemetry and orbital data from the Satellite Control Centre, ancillary data from laboratories.
- Data processing: payload telemetries and ancillary data are processed to generate the mission products. These products are organised according to their level of processing. Three to four levels can be considered.
- Data delivery: all the data acquired or produced by the Payload Data Centre can be dispatched to the users. Delivery rules are defined by a Scientific Committee.
- Expertises: a toolset can be provided to the mission user to check the correct operations of the instrument and the quality of the mission products.
- Data configuration management: all the data must be stored and securely saved as they are the mission memory.

A Payload Data Centre communicates with various entities as the Satellite Control Centre, the Principal Investigator (PI) and Co-Investigator (CoI) laboratories, some exogenous centres and the mission users. Internal interfaces between Payload Data Centre sub-systems are also possible when it is spread out over different facilities.

## III. Presentation of the PICARD Payload Data Centre

### A. Responsibilities

The PICARD Payload Data Centre is an operational data processing centre, which will be used during all mission phases. It will take charge of the programming and the monitoring of the payload (SODISM, SOVAP, PREMOS), the storage, the processing, the acquisition and the distribution of the data related to the mission. All these tasks can be triggered either automatically or manually. Figure 4 shows the functional structure of the PICARD Payload Data Centre. Functionalities and roles of external entities are described afterwards.

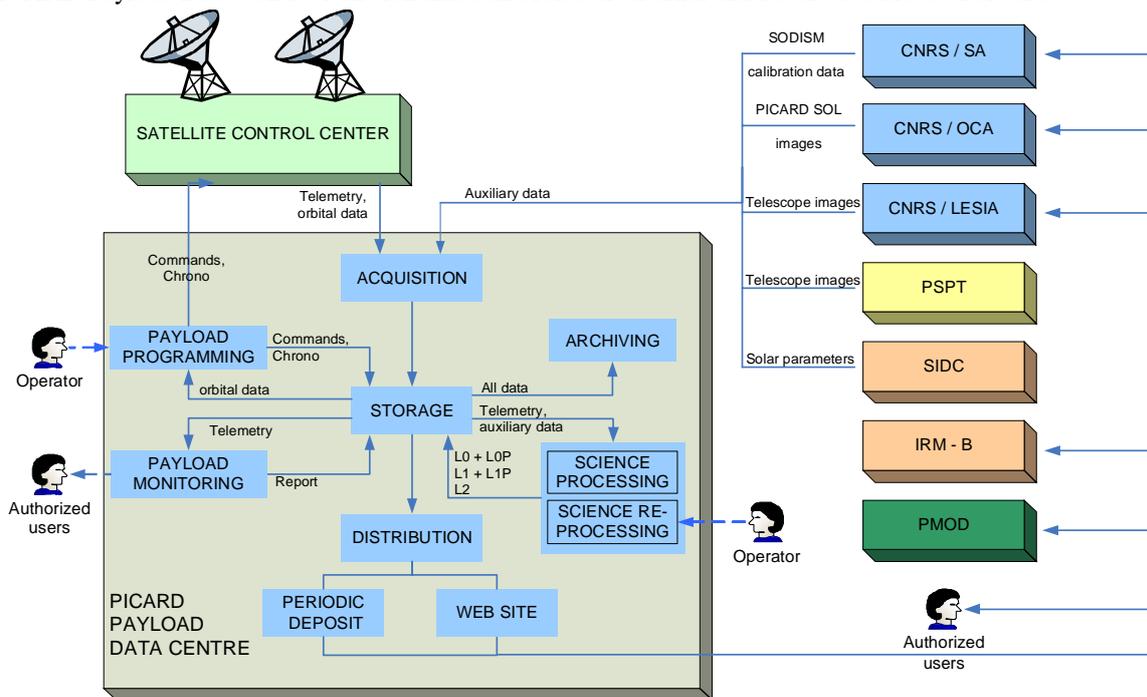

**Figure 4. PICARD Payload Data Centre functional structure.**

The PICARD Payload Data Centre stakeholders are :
- The operator: it is the person in charge of the correct progress of the operations. He periodically controls the acquisition of the data, he performs twice a week the payload programming plan and he carries out all the data management operations (processing, storage, distribution, web site updating). The operator takes each time into account the commands emanated from the mission committees (Scientific and Operation Committee).
- The Operation Committee: once a week, the Operation Committee meets to analyse the activity related to the mission and to decide about the next satellite operations. This group takes charge of the platform and the payload. It is composed of CNES engineers, scientist representatives and the operator.
- The Scientific Committee: this group assists the mission PI in the elaboration of the mission requirements, the content of the scientific products, the data access policy, the scientific program related to the mission and the payload programming scenario. It is composed of the PICARD mission PI, the instrument PI and laboratories representatives.

The PICARD Payload Data Centre lifespan is 6 years: 6 months before launch, 6 months during the acceptance test phase, 3 years of nominal mission and 2 years of post-mission operations. The responsibility during each step is described below.

- Technical and operational on-ground tests: this phase takes place before the satellite launch. All the processes will be tested manually and automatically in order to check all the triggering possible cases. The most tested process will be the payload programming since it is the most critical function.
- In-flight acceptance tests of the satellite (payload and platform): this phase corresponds to the very first weeks of the mission. The payload will be switch on and the first in-orbit tests will be performed. During this phase, all the processes are triggered manually in order to send many commands with various configurations. This phase is planned to take 3 months.
- Acceptance tests of the scientific processing: this phase follows the previous one and will take 2 months. During this period, various versions of the scientific algorithms implemented into the PICARD Payload Data Centre will be tested. All the processing sequences will be individually tested. The payload programming plan will be automatically elaborated.
- Nominal mission: this phase is intended for 3 years however the total duration is not limited. All the processes will be automatically triggered. The PICARD Payload Data Centre should be able to carry out a programming plan, to collect all the data necessary to the data processing sequences, to monitor the payload housekeeping data, to elaborate the mission products, to store all the data and to distribute them according to a scientific policy established by the PICARD Scientific Committee. During this phase, the unavailability of the PICARD Payload Data Centre must be less than five days.
- Satellite removal: the PICARD Payload Data Centre must continue to operate for two years after the nominal phase of the mission. During this period, it takes on the re-processing wanted by the Scientific Committee, transfers all the data to a specific centre that will store them for a long duration period.

### B. Hardware architecture

The PICARD Payload Data Centre hardware architecture is as follows :
- A main server hosting the processing centre.
- A storing system of 24 TBytes dedicated to store all the mission information. This system is accessed by the computing server through two redundant Gbits ethernet switches.
- A broadcasting/internet server hosting the distribution centre.
- An « Uninteruptible Power Supply » that enables to better manage the supply breaking.

The hardware is integrated into a high rack. A back-up hardware is coupled with this operational hardware in order to allow the continuity of services in case of major damage to the operational configuration. PICARD Payload Data Centre services are operated and managed from two dedicated consoles. This management is done through the local network with a specifically developed interface.

### C. Software architecture

Being a scientific mission, the PICARD requirements are likely to evolve during the mission development and during the operations. The PICARD Payload Data Centre software architecture must be flexible. It has been considered that common tasks as storage and database management can remain unchanged while the mission specific tasks (programming, processing, acquisition, distribution, control) must be configurable. The architecture is composed of a structure hosting the common tasks and of a library of independent services (see Figure 5).

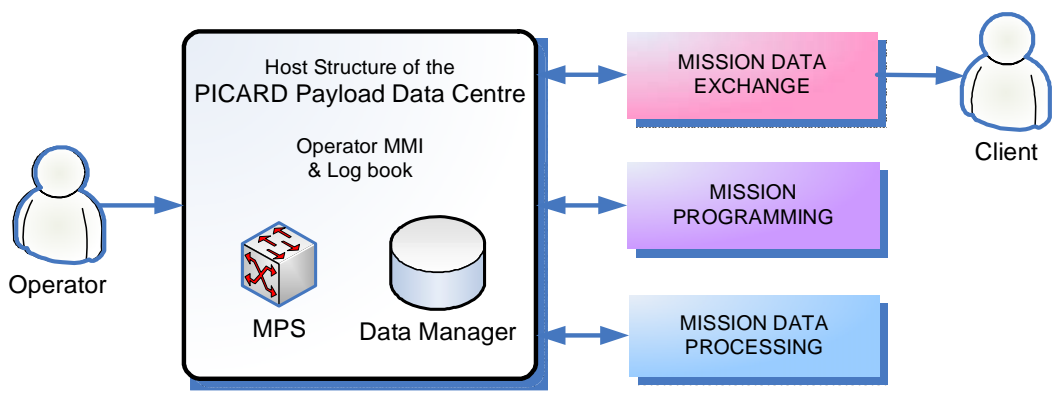

**Figure 5. Architecture of the PICARD Payload Data Centre**

The host structure is composed of a main Man-Machine Interface (MMI), a Data Manager and a Mission Planning and Scheduling System (MPS). The library of service contains all the executable files and scripts necessary to the exchange, programming and processing functions.

*C.1. The host structure MMI*

It is the operator main interface. It informs about the current state of the triggered services, allows the operator interacting with their execution or the operator triggering a service. The host structure MMI is linked to a log book where messages provided by the different services or by the host structure itself are listed. From the log book, it is possible to explore all the previous messages by different criteria: time period, message type (normal, warning, error) and message content. The PICARD Payload Data Centre can easily be configured in three languages: French, English and Dutch.

*C.2. The data manager*

The data manager is responsible for the database access. It takes charge for the management of the stored data for all the services. It is used by the processing server and by the distribution server. A storage system manages a direct access to the data catalogue and a back-up recording on magnetic tape. During the mission, data are compressed and recorded on hard drives, which capacity is about 24 TBytes. The back up system uses magnetic tapes, which volume is about 400 Giga Bytes. The PICARD Payload Data Centre preserves all the collected and produced data. This is mandatory for allowing data re-processing.

| Level | Daily data | Daily compressed data | Mission compressed data |
|---|---|---|---|
| Raw | 0.300 GB | 0.300 GB | 330 GB |
| L0 | 7.50 GB | 1 GB | 1100 GB |
| L1 | 14 GB | 1 GB | 1100 GB |
| L2 | 0.02 GB | 0.02 GB | 22 GB |
| **Total** | **21.82 GB** | **2.32 GB** | **2 552 GB** |

**Table 1. Data volume of the PICARD mission (predicted values).**

*C.3. The mission planning and scheduling system (MPS)*

It is used to manage the service triggering. The operator can modify the triggering criteria: immediate, delayed or periodic. A request is made to a sequencer that handles the execution of the service. The operator can also interrupts the service. The MPS is able to manage parallel processing.

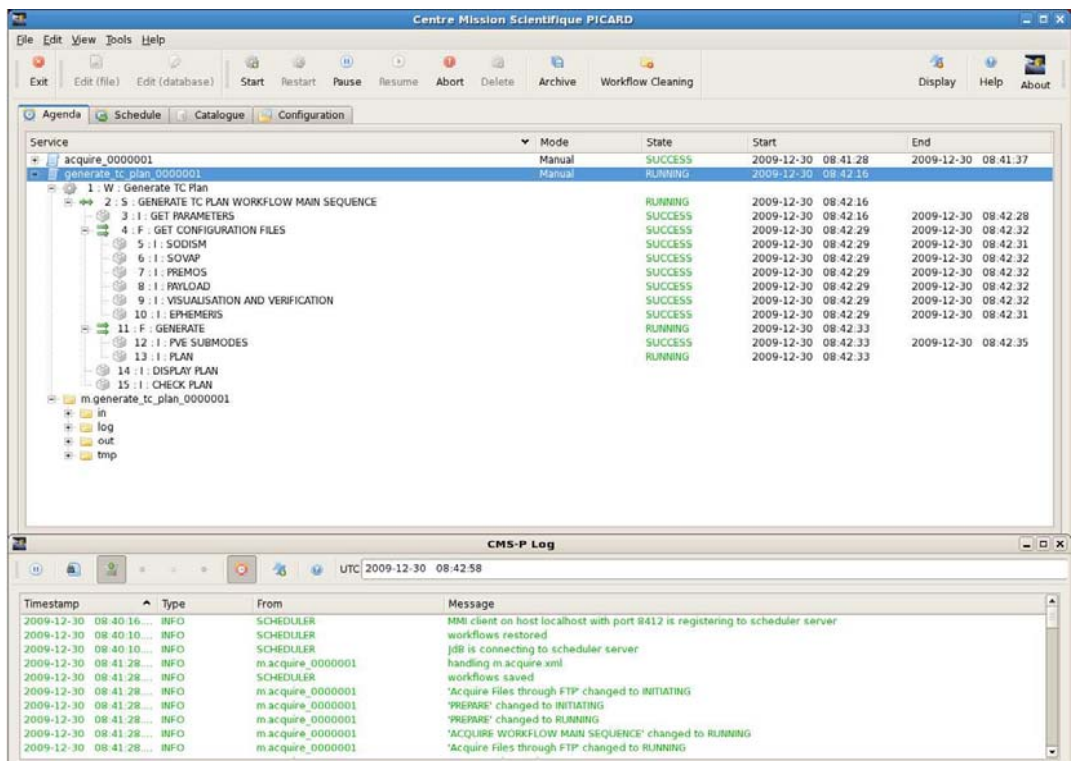

**Figure 6. Main interface screenshot showing the planning system (top window) and the log book (bottom window).**

*C.4. The library of services*

It is composed of all the executable programs and scripts that can be applied on the PICARD data. The services are organised in three groups.
- Mission Data Exchange services (see D): scripts for collecting or diffusing data files by File Transfer Protocol (FTP). Diffusion of data is also carried out by Hypertext Transfer Protocol (HTTP). Each type of exchange is defined by a script that indicates the connexion parameters and the data files to be collected or distributed. Each parameter is modifiable. The data exchange services regroup an automatic process (scheduled exchange) and an on-request process. The latter uses a web site for allowing a distant consultation of the database. The on-request distribution also differs by the possibility to directly visualize certain type of data.
- Mission Programming service (see E): executable programs or scripts developed by the industry to produce the payload programming plan. This service uses a pre-defined programming scenario to elaborate a solution of command plan. This service has its own MMI to allow the operator modifying the plan.
- Mission Data Processing services (see F): executable programs or scripts applied to the PICARD data to create the mission products. Software components of the scientific processing sequences are developed by the laboratories involved in the mission. The sequences are afterwards integrated by the industry in charge of the PICARD Payload Data Centre realisation. The other software components, as processing of the housekeeping data, are provided by the industry.

## D. Data exchange service

Figure 4 shows clients of this service. There are the Satellite Control Centre, the laboratories involved in the mission and the authorised users (public users, external laboratories). Interface with each entities is described below. Data to be exchanged are not critical, only the programming plan must be signed to ensure the integrity of the file during the transfer.
- Acquisition: the PICARD Payload Data Centre collects only new data from external servers and data, which name is known by the database. A version number allows differentiating data, which name is similar.
- Diffusion: all the data available in the database (collected or produced data) can be distributed to the authorised clients. A part of these data is periodically provided by the distribution server. Authorised clients can simply collect them through an automatic FTP process. Another way consists in searching data directly into the database using Internet. This method is currently under development. The objective is to allow the authorised clients using various criteria (instrument, data level or type, header parameters range) to found the requested data.

*D.1. Interface with the Satellite Control Centre*

It is the main data provider of the PICARD Payload Data Centre since it delivers all the payload telemetries and the auxiliary mission data (satellite orbit and attitude, orbital events predictions). The Satellite Control Centre uses the CNES secured server to exchange data with the PICARD Payload Data Centre. The volume is small (300 MBytes/day for the telemetry, several KBytes for the programming plan). Furthermore, there is no need for real time availability. All the data are thus transferred via Internet. The following data are provided by the Satellite Control Centre. They remain available for 10 days.
- Scientific telemetry: these files contains the scientific payload data split into packets using the format recommended by the Consultative Comittee on Space Data System (CCSDS).
- Housekeeping telemetry: these files contains the housekeeping data of the platform and of the payload. They are split into CCSDS format packets.
- Predicted orbital events: this file provides all the events occurring along the orbit for a 3-week duration. Each event is described by a date, a class, the satellite orbital position and a commentary. Events concern the orbit (ascending/descending node, day/night transition, polar zones, atmospheric occultation zone), the Earth terminals (entrance/exit time, elevation), the satellite (guidance system, propulsion events) and the PICARD mission. The latter are used to program the mission mode (see F).
- Predicted satellite orbit: these files contain the estimated position and velocity of the satellite in inertial (J2000) and terrestrial frame (WGS84) for a 3-week duration. It is updated each day.
- Determinated satellite orbit: these files contain the position and velocity of the satellite in inertial (J2000) and terrestrial frame (WGS84). It is calculated each days over the previous day. The position and velocity are determined by the Doppler method (no Global Positioning System, no laser reflector).
- Determinated satellite attitude: this file contains the satellite quaternion. It is calculated each day, over last 24 h, from the star sensors measurements (4 Hz sampling).

The PICARD Payload Data Centre provides the following data.
- The programming plan: it is processed twice a week. A full description of the programming process is

given in E. This file is considered as sensitive. It is indeed sent by the Satellite Control Centre without preliminary check. Even if no damage can be performed, it is necessary to valid the integrity of the file to avoid any disturbance during the mission. The control is based on the exchange of numerical key between the Satellite Control Centre and the Payload Data Centre.
- The programming description: it is the description in text of the programming plan content. It provides the date and parameters values of each command. The file is produced by the PICARD Payload Data Centre and put at disposal of the operators to investigate the programming when problems occur during the mission.

*D.2.      Interface with the CNRS/SA*

In addition to its scientific activity as PI of the PICARD mission, the Service d'Aéronomie of the CNRS is a main contributor to the PICARD Payload Data Centre. Before the mission starts, it handles the developments of all the scientific processing (coding and configuration requirements, delivery planning, documentation) to be implemented into the PICARD Payload Data Centre system. It provides also the reference data necessary to validate the scientific data processing sequences. During the mission, it produces the SODISM calibration data. These data will be transferred to the PICARD Payload Data Centre to be used within the level 1 processing sequences. The calculation uses the flat-field recorded by the instrument and recovered each day. The laboratory provides also data of the SOLar SPECtrum (SOLSPEC) experiment that characterising the solar spectral energy (with respect to the wavelength).

*D.3.      Interface with the CNRS/OCA*

The Observatoire de la Côte d'Azur (CNRS/OCA) is the PICARD SOL data provider. A similar model (SODISM II) of the space SODISM telescope will be installed on the Calern shelf. PICARD Payload Data Centre will collect each day some solar limb at 393, 535, 607 and 782 nm, some full sun image at 393 nm and data characterising the atmosphere at the measurement date (Irbah [14]). The number of available data depends on the atmospheric conditions. The PICARD Payload Data Centre does not process the collected SODISM II measurements. The CNRS/OCA laboratory collects from the programming description of SODISM.

*D.4.      Interface with the SIDC*

The Solar Influence Data Analysis Centre (SIDC) is responsible for providing the data characterising the solar activity. These data will be used by processing sequences of the PICARD Payload Data Centre.

*D.5.      Interface with on-ground observatories*

The PICARD Payload Data Centre collects each day images from on-ground observatories:
- Laboratoire d'Etudes et d'Instrumentation en Astrophysique (LESIA): 2 full solar images at 393.4 nm and 393.3 nm
- Precision Solar Photometric Telescope (PSPT): 4 full solar images at 393.4 nm, 393.6 nm, 409.4 nm, 607.1 nm.

These data are not processed but only distributed to help in the data interpretation. As the SODISM II measurements, the availability of the LESIA and PSPT images depends on the atmospheric conditions.

*D.6.      Interfaces with the authorised users*

All the collected and produced data can be potentially distributed. The Scientific Committee establishes the distribution restriction of the PICARD data. Scientific mission products are performed each day (see F). Products of level 0 are available the day after the telemetry acquisition. Products of level 1 and 2 are available the day after the level 0 creation. Housekeeping data are processed once a day. A report on the payload status is available each morning.

**E. Payload programming service**

The PICARD Payload Data Centre takes charge for the nominal programming of the payload that is carried out twice a week by the operator. It uses constraints given by the Operation Committee, the Scientific Committee and the nominal programming scenario that is established and validated before launch. The Satellite Control Centre takes charge for switching on and off procedures of the three instruments. The PICARD Payload Data Centre is responsible for the mission programming that is the determination of each measurement dates.

*E.1.      SODISM programming*

SODISM is a telescope using a 2048x2048 pixels CCD (Thuillier [6]). The relative accuracy of the measurement is about $10^{-3}$ arc-second. It is the most complicated instrument to be operated since there are various different types of measurements. Each type is defined by a 2 bytes word that indicates the wavelength (dark current, 215, 393, 535, 607 and 782 nm), the image size (narrow/wide limb, full/windowed images), the on board processing (software compression), etc. A table of about 65 different types has been created. Only a small part of them are used during the nominal mission. The other types are used during the In flight Acceptance Test Phase or during investigation phases. Figure 7 shows the various image types of the SODISM telescope.

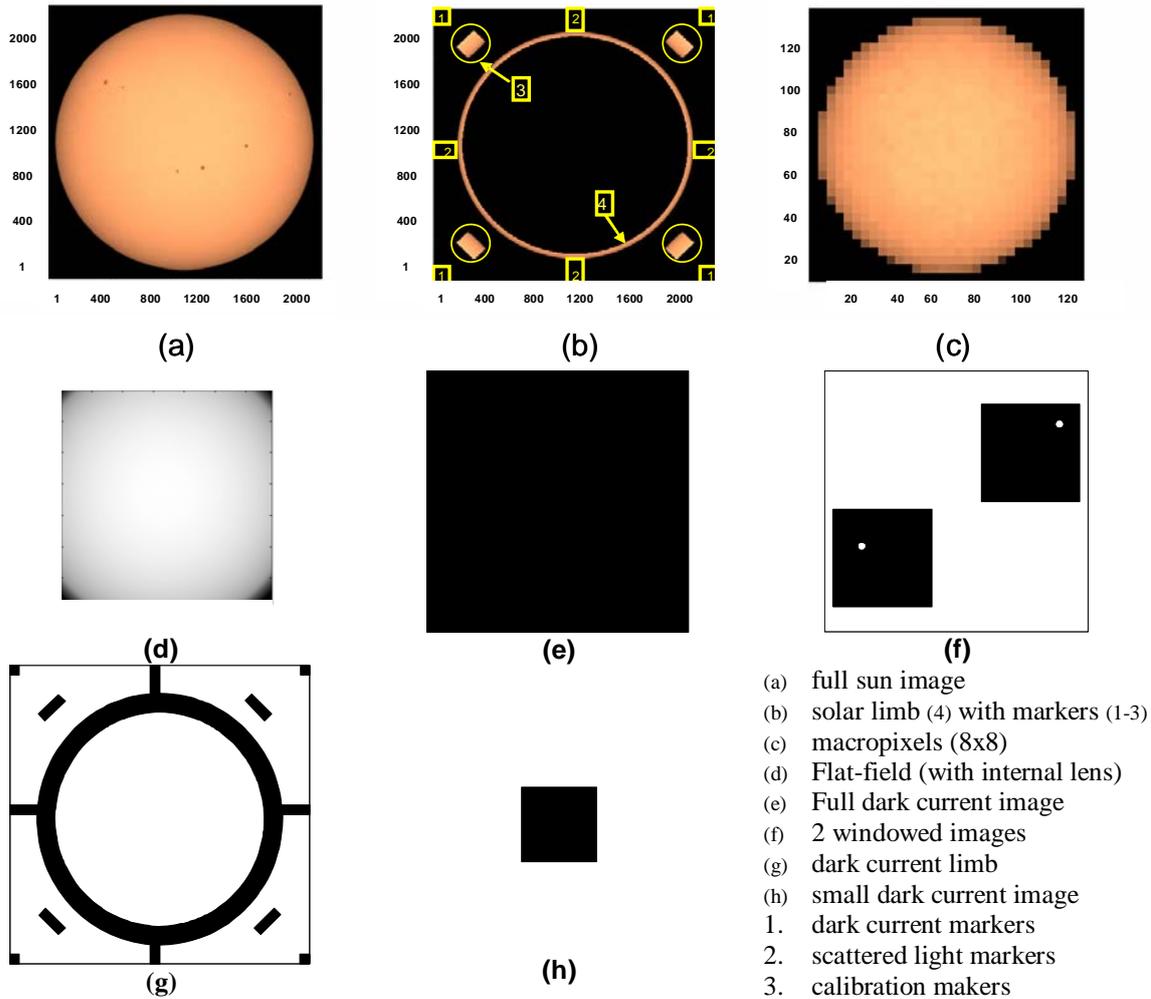

**Figure 7. Image types of the SODISM telescope.**

Given the limited telemetry volume per day, two types of compression are used on board:
- A loss-less compression provided by CNES named PICARD Loss Less Cruncher (PLLC)
- A compression with loss at a rate of 16, which has been design by Y. Langevin (Langevin [15])

This rate has been chosen to prevent artefact effect on sunspots position used to retrieve the sun differential rotation. The nominal configuration uses the following image types:
- Limbs: 22 or 40 pixels width, compressed using PLLC,
- Full images: 2048 x 2048 pixels, compressed using the Langevin algorithm,
- Macro-pixelised images: average over 8x8 pixels, compressed using PLLC algorithm,
- Windowed images : 768 x 768 pixels, compressed using PLLC algorithm,
- Small images : 256 x 256 pixels, not compressed

Each image type can be obtained whatever the wavelength filter is. When the shutter of the instrument is maintained closed, a dark current image is recorded. It must be outlined that each measurement provides a full sun image that is processed on board. One measurement can also provide two images: a full sun image and the corresponding solar limb.

The SODISM instrument can be operated in two modes, the measurement mode and the investigation mode. The programming scenario always uses the measurement mode. The investigation mode is a mode used for specific check and is manually operated. The measurement modes are:
- **Nominal mission mode**: it is the scenario used by default, when there are no satellite manoeuvres, no calibration phase and no physical events such as Earth's eclipse. This scenario consists in taking a macro-pixelised image of the Sun (535 nm) each minute, a narrow limb of the Sun (535nm) each two minutes, two wide limbs of the Sun for each wavelength (215, 393, 535, 607 and 782 nm) per orbit and two full images of the Sun (one image with 393 nm filter and one image with another filter) per orbit. In addition to these solar images, calibration pictures have to be taken: four wide limbs of dark current per orbit and before the solar limbs measurements, one full image of dark current and one full image of flat-field (using a divergent lens) per day. Furthermore, the solar limbs and full images must be

- **Optical distortion mission mode**: this is a calibration scenario to be operated once a month. The satellite rotates twelve times around the solar direction axis by step of 30 arcdegree. Before each rotation, ten wide limbs at each wavelength are recorded (215, 393, 535, 607 and 782 nm). Theses measurements will be used on ground to characterise the optics and CCD. A modified scenario has been elaborated for the case of an eclipse. This phase lasts a maximum of twelves orbits.
- **Stellar mission mode**: this is a calibration mode too. It is foreseen to be operated four times a year. The satellite is rotating to be pointed toward the barycentre of two stars having an angular distance of about 30 arcminutes. This operation takes about 65 minutes including two spacecraft rotations. After rotating, twenty windowed images are recorded. They will be used to characterise the angular/pixel relationship.
- **Absorption mission mode**: this mode is enabled when the instrument sighting axis is crossing the Earth atmosphere at an altitude lower than 40 km. In this condition, the nominal mode cannot be used due to the disturbing effect generated by the atmosphere. A sequence of solar wide limbs at a selected wavelength is taken. This mode might not occur very often, less than 150 orbits per year.
- **Night mission mode**: this mode is enabled when the Sun is hidden by the Earth. Eclipses are intended to occur once a year during three months. The maximum duration of the eclipse is 20 minutes per orbit. During that period, dark current images are taken. When the satellite leaves the shadow, a sequence of solar limbs is taken during a configurable period. These measurements will be used to characterise the thermal behaviour of the instrument and validate the model, which describe its optical characteristics with the temperature.
- **Dark current mission mode**: this is a specific mode to be used during the In flight Acceptance Test Phase of the mission for identifying the particles precipitation zones. It consists of taking a sequence of dark current small images or full images all along the orbit and during a few days.
- **Investigation mode**: this mode is enabled when the operator has to program measurements that are out of the of the previous scenarii.

Table 2 indicates the number of SODISM images per scenario and per day.

| IMAGE TYPE | MISSION MODE | IMG / DAY |
|---|---|---|
| Macropixel 8x8 pixels, 535 nm | Nominal | 1235 |
| | Night | 294 |
| Limb, 20 pixels, 535 nm | Nominal | 720 |
| | Night | 294 |
| Limb 40 pixels, dark current | Nominal | 58 |
| | Optical distorsion | 60 |
| | Night | 210 |
| Limbs 40 pixels, 215, 393, 535, 607, 782 nm | Nominal | 145 |
| | Optical distorsion | 600 |
| | Absorption | 135 |
| Full image, 2048 x 2048 pixels, 393 nm | Nominal | 14 |
| Full image, 2048 x 2048 pixels, 215, 535, 607, 782nm | Nominal | 14 |
| Full image, 2048 x 2048 pixels, dark current | Nominal | 1 |
| | Dark current1 | 100 |
| Full Flat-field image, 2048 x 2048 pixels, 215, 535, 607, 782nm | Nominal | 1 |
| Windowed images, 768x768 pixels | Stellar | 20 |
| Small images, 256x256 pixels | Dark current2 | 1440 |
| **TOTAL (total number of images per scenario and per day)** | | |
| Nominal | | 2188 |
| Optical distorsion | | 660 |
| Night | | 798 |
| Absorption | | 135 |
| Dark current1 | | 100 |
| Dark current2 | | 1440 |

Table 2. Number of the images foreseen with the SODISM mission scenario. Notes that the total number of images per day can be superior to 1440 since one measurement can provide two images.

E.2. *SOVAP programming*

SOVAP measures the total solar irradiance among a sequence of radiometric states, which is selected by command (Dewitte [18]). The duration of each state is 90 seconds. The same sequence is continuously used for the solar observations. The SOVAP data are daily stored. They are composed of several 90-seconds frames containing science and housekeeping data. The use of these frames allows calculation of the absolute and relative total solar irradiance with a sampling time of respectively 3 minutes for the total solar irradiance, which is later sampled at 10 s using the bolometric measurement. A specific sequence is used once a month for the calibration of the instrument.

There is only one mission mode for the SOVAP instrument. The instrument programming is not orbital event dependent. There is no synchronisation requirement to be managed by the PICARD Payload Data Centre when calculating the SOVAP programming.

### E.3. PREMOS programming

PREMOS measures the spectral solar irradiance with three filter radiometers (210, 266, 535, 607 and 782 nm) and the total irradiance with 2 radiometers (Fröhlich [19]).

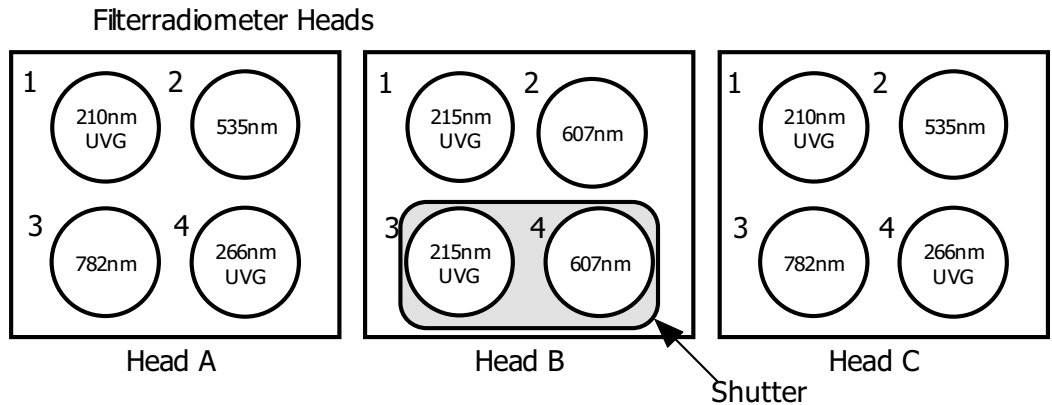

**Figure 8. Three filter radiometers with Detector- and Filter- Type**

The spectral and total solar irradiance measurements are carried out continuously with the same heads and at precise interval time with a duplication of measuring channels of the sunphotometers and radiometers. Other useful data for calibration and instrument control are measured and inserted in the science telemetry. These housekeeping data are stored in five daily products at the PICARD Payload Data Centre while the science data are recorded within two files. Various constraints have to be taken into account to calculate the instrument programming.

- Calibration phases: solar measurements can be performed by several sunphotometers as redundancies are foreseen. A photometer and a radiometer work quasi continuously except during the calibration phases for which the sunphotometerss are operated. The calibration phases are periodically executed. The frequency and the duration are selected during the In flight Acceptance Phase of the mission.
- Instrumental modes: there are two instrumental modes, the nominal mode and the night mode. The latter corresponds to a configuration where all the shutters are closed. This must be activated when the instrument is not lit by the Sun (eclipse or stellar pointing). The PREMOS programming is thus orbital event dependent.
- Payload synchronisation: the PICARD Payload Data Centre and the embedded software have to manage a synchronisation between the PREMOS and the SODISM measurements when the latter takes full images of the Sun (excepted for 268 nm as this wavelength is not available with SODISM).

### E.4. PICARD Payload Data Centre programming tool

From the above scientific scenarii, a specific algorithm has been developed by CNES and industry to automatically elaborate a programming plan satisfying all the constraints This algorithm uses the orbital events prediction to identify when each mission mode of the three instruments have to be enabled. The orbital events data are automatically provided by CNES each morning (see D.1). The programming plan is deduced from this calculation. At the end of the process, a graphical view of the programming plan is shown to the operator (Figure 9). The latter can modify measurements and commands configuration. Another solution consists in launching the graphical tool to directly build the programming plan. The operator can either start from a previous plan and update it or start from an empty sheet. This graphical view of the programming plan is a very powerful tool. It shows on a same view the orbital events, the measurements type of each instruments, the satellite orbit and the measurement modes. Simple clicks or copy/paste allow the operator modifying easily the programming plan.

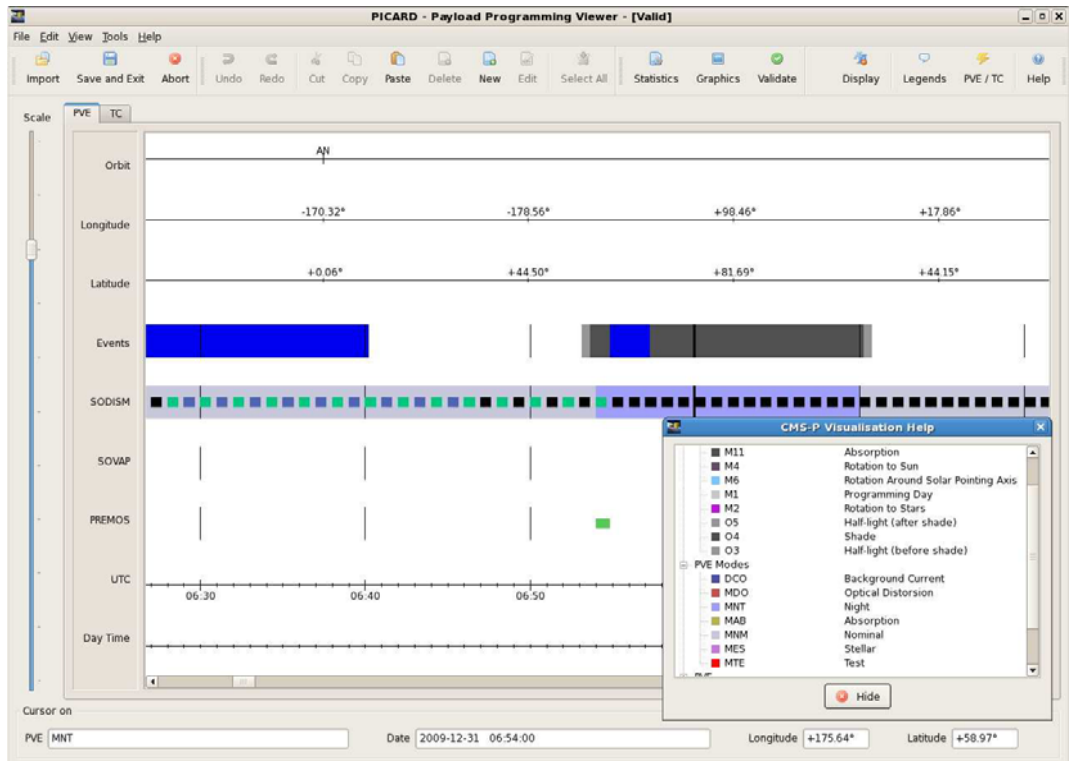

**Figure 9. Screenshot of the programming plan visual tool. The operator can modify the dates and the type of the measurements. A colour code allows identifying the various measurement modes or orbital events.**

The programming plan must be checked before being sent to the Satellite Control Centre. The requirements to be satisfied concern the measurement sequences and the command sequences. They are given by the Scientific and Operation Committees. The nominal programming is carried out for a 7-day duration. A time offset can be inserted between the day where the programming plan is performed and the day for which it must be applicable. Furthermore, a back-up period of three days is considered for each plan.

**F.  Scientific data processing service**

CNRS/SA handles the development of the scientific algorithms to be implemented into the PICARD Payload Data Centre. CNRS/SA is the CNES correspondent with the laboratories developing the scientific software. Thus, the source code are issued to the CNRS/SA that verifies the code standard requirements: all algorithms need to be developed according to the same requirements as coding languages (C or Fortran 90), coding rules, configuration description file, auto-test data, compilation and assembling procedures. Industry takes charge for the integration of these softwares and the creation of the various processing sequences. Only executable codes with their test data are provided to the industry. The PICARD mission products and scientific processing are presented below. First, we have to know how the data are processed onboard to understand algorithms that will be integrated in the PICARD Payload Data Centre system to create level 0 products (see definition below). Onboard data processing concerns mainly the SODISM images. This is due to the large volume of data generated by the instrument relatively to the allowed rate of transmission.

*F.1.   Onboard data processing*

The CCD camera of SODISM provides images of size 2160x2153 pixels (the total size is bigger than 2048 x 2048 due to supplementary colons and rows). A mask is applied to the full image to generate a limb. This mask is built to also acquire parts of the full CCD image that will be used as makers. A transformation of the limb image is then performed onboard to keep only the non-zero pixels. They are projected on the y-axis of symmetry of the image. This is callas the "Os transformation" (Figure 9).

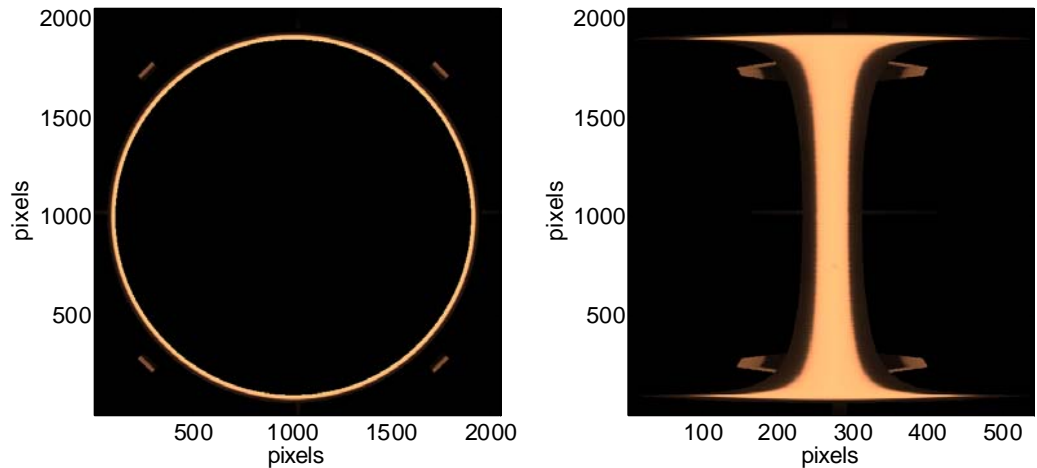
**Figure 10. Example of the "Os transformation" of a solar limb (with its markers)**

Another onboard processing concerns the creation of macro-pixelised images. They are calculated by convolving a full image with a 8x8 pixels spatial window. The last onboard processing consists in the compression of the data before transmission using either the PLLC or Langevin algorithm (see E.1). After these processing, the onboard software creates the telemetry packages with the CCSDS format. It adds also some housekeeping data and image parameters within the package.

*F.2. SODISM data processing*

For creating Level 0 (L0) products, data of the image are extracted from the CCSDS packets. This binary stream can afterwards be decompressed if needed. For limb image types, the inverse "Os transformation" is performed with the same mask to the one used on board. Makers of the limb image are extracted to create daily L0 products. No further data processing is required for creating the full images products (macro-pixels, dark signal, flat field and windowed images). The creation of L0 products requires five processing sequences involving eight software components. The SODISM nominal mission mode is intended to produce 2 188 N0 products per day, that is almost 2 500 000 images for the full mission. A quick analysis is also carrying out on the limb's markers and on the dark current full image. It is to compute the mean and standard deviation of pixels intensity but also to count and to locate the bad pixels in the image. L0 Quick Look requires three processing sequences with five software components.

The Level 1 (L1) processing sequences concern the calibration of the L0 products (scientific and housekeeping data) with adding new information in their header to create self-consistent L1 products. They are also perform to create daily products containing the auxiliary images presents in the limb images. Theses products are used to compute instrument scale factors (Assus 13). For the helioseismology mission objective, others processing sequences are implemented to monitor the intensity of the solar limb and the photometry of the macro-pixelised images. They also compute a $\ell$-$\nu$ diagram diagnostic (Corbard [14]). The creation of the L1 products requires five processing sequences and nine software components. The number of L1 products per day and for the full mission is similar to the L0 products.

Concerning the quick look processing sequence, one L1 Quick Look product contains the power spectrum of the temporal fluctuations of the intensity of the limb and of the macro-pixelised images. The intensity is corrected from Satellite-Sun distance. The temporal intensities of macro-pixelised images projected over spherical harmonics basis (YLM masks) are also computed to create a second Quick Look product. The last quick look is an intensity analysis of full image recorded at several wavelengths. The obtained solar spectral irradiance is compared with the same value given by the PREMOS instrument or from the other space mission as SOLSPEC. The creation of the L1 Quick Look products requires two processing sequences and three software components.

The level 2 (L2) processing sequences are to compute the mean solar radius value from each limb image and its daily variations. The accuracy of measurements and noise are also estimated. L2 products that are created are unique for the entire duration of the PICARD mission. Two processing sequences involving three software components are needed for creating L2 products.

*F.3. SOVAP data processing*

Level 1 processing sequences of SOVAP control the data quality of the daily L0 products. They are calibrated afterward and converted in electrical units. A product with a specific format is then created in order to be easily viewed with appropriate quick look tools. Three software components are needed for this processing. Level 2 processing requires one software component. It computes the total solar irradiance at ten seconds and at three minutes sampling times. The daily mean values are also computed as well as the aging corrected values.

*F.4.    PREMOS data processing*

The first level of PREMOS processing is to calibrate all data (science and housekeeping) in electrical units and temperature. Some housekeeping data are also taken into account to correct science data from thermal effects during calibration. All L0 products are considered to create L1, which will be input to the level two processing sequences. They compute the total solar irradiance at two minutes sampling times but also average values over one hour and one day. These values are corrected from orbital parameters. The same calculations are made for solar spectral irradiance measured at the PREMOS wavelengths. The level 2 processing sequence compares the measured values of the spectral irradiance to a specific one and to the total solar irradiance recorded at the same sampling times. Total solar irradiance obtained with SOVAP and PREMOS are also compared. All these processing sequences are made using 6 software components.

*F.5.    Product name*

The name of each product is created using acronym for the mission name, the instrument name, the product level and some data type identifier (mission mode, measurement type and sub-type). The date of measurement and the product version are also mentioned. The header of each product contains more information of the instrument housekeeping data and the processing sequence.

*F.6.    Payload monitoring service*

The processing of the payload housekeeping data consists in comparing the values to a reference threshold or checking the status of a parameter. All the housekeeping data of the payload (and some data of the satellite platform) are collected by the PICARD Payload Data Centre from the CNES secured FTP server. The data are calibrated, organised by type and stored within the database. A control of some parameters is automatically performed each day. This service provides a bulletin where only alarms are mentioned. This bulletin is automatically set at the disposal of laboratories onto the FTP server of the PICARD Payload Data Centre.

## IV.    Exploitation of the PICARD Payload Data Centre

### A. Presentation of the B-USOC

The main role of the Belgian User Support and Operation Centre is to promote space research programmes and flight opportunities to Belgian scientific community in universities and federal institutions. Subsequently, it provides support to scientists concerning the definition, the development and the operation of their experiments conducted in the different space research fields: microgravity, earth observations, space sciences and space technology. The centre supplies to scientists or users a set of services and resources to prepare, perform, monitor and analyse experiments on board a Facility on the ISS but also on board of Satellite platforms.  There will be three possibilities for the user/scientist:
1. Direct use of the hardware available at B-USOC,
2. Use of his/her own hardware that must comply with the interfaces within B-USOC,
3. Use of his own hardware according to the proposed standardized communication facilities.

B-USOC also provides services equivalent to the ones provided to the ESA for the space missions sustained by bilateral agreements. In this specific context, the centre has the overall responsibility for the implementation, preparation and execution of operations for the future PICARD Payload Data Centre.

### B. PICARD Payload Data Centre Operations

Most of the activities are automatically handled by the PICARD Payload Data Centre. Nevertheless three operators will be successively present during working hours. Their work mostly consists in monitoring the processes, but they also handle any unusual activity (non-routine operation of the instruments, update of scientific algorithms, re-processing of data with new calibration matrix, fixing of any detected anomaly). They also have a communication role, interacting with the scientists as well as with the control centre, and updating the website of the mission. Among others, operators of the PICARD Payload Data Centre have regular contacts with the Operation Committee and with the Scientific Committee. The Operation Committee determines the routine and non-routine operations that will be performed with the three instruments.

The hardware configuration has been described elsewhere. As a reminder, it is composed of a nominal and a redundant configuration, the redundant one being used in case the nominal one crashes. For security reasons, accesses to the PICARD Payload Data Centre are restricted by both physical means (magnetic card are needed to access the PICARD Payload Data Centre room) and software means (use of passwords).

The B-USOC has been involved very early in the PICARD mission preparation. One of the operators is a member of the CNES integrated team, which supervises the payload on-board and ground-segment developments (with, among others, the development of the PICARD Payload Data Centre).

The goal is that the operators have a deep knowledge of the instruments they are operating and benefit from good diagnosis capabilities. Once the Payload Data Centre will be tested and validated, it will be delivered to the B-USOC where it will be installed. The CNES will then hand over to the B-USOC for the PICARD Payload Data Centre exploitation.

## V. Conclusion

Given the excellent synergies between all the team involved in the Payload Data Centre project (laboratories, agencies and companies), the latter is going to provide all the functionalities required to operate the PICARD scientific experiment. It allows laboratories programming theirs instruments, collecting daily products of the mission, controlling the status of theirs instruments, asking for re-processing and storing the mission data in a unique centre. Furthermore, the coming developments will provide on-line functions that will offer users a better data access and a better visibility of the mission .

## VI. Acknowledgments

Authors would like to thanks all the team involved in the financing and the development of the PICARD Payload Data Centre for their confidence, support and motivated contribution: BELSPO, BUSOC, CNES, CNRS, CNRS/SA, CNRS/OCA, CNRS/IAS, CNRS/LESIA, ESA, IMCCE, IRM-B, PMOD, PSPT, SIDC, SPACEBEL.